\documentclass[aps,pre,reprint,groupedaddress]{revtex4-2}
\usepackage{graphicx}
 \usepackage{amsmath}
\usepackage{amssymb}
\usepackage[font=small,labelfont=bf]{caption}
\usepackage{amsfonts}

\usepackage{subfig}
\usepackage[utf8]{inputenc}
\usepackage{dcolumn}
\usepackage{seqsplit}
\usepackage{natbib}
\usepackage{relsize}

\date{\today}
\newcommand{\be}{\begin{equation}}
	\newcommand{\bea}{\begin{eqnarray}}
		\newcommand{\bc}{\begin{center}}            
			\newcommand{\ee}{\end{equation}}
		\newcommand{\eea}{\end{eqnarray}}
	\newcommand{\ec}{\end{center}}

\newcommand{\baa}{\begin{eqnarray*}}
	\newcommand{\eaa}{\end{eqnarray*}}
\begin{document}
	\title{Thermoelectric generator in endoreversible approximation: \\
	the effect of heat-transfer law under finite physical dimensions constraint 
}
	\author{Jasleen Kaur}
	\email[e-mail: ]{jasleenkaur@iisermohali.ac.in}
	\author{Ramandeep S. Johal} 
	\email[e-mail: ]{rsjohal@iisermohali.ac.in}
	\affiliation{Department of Physical Sciences,\\
		Indian Institute of Science Education and Research Mohali,\\
		Sector 81, S.A.S. Nagar,\\
		Manauli PO 140306, Punjab, India.}
	\author{Michel Feidt}
	\email[e-mail: ]{michel.feidt@univ-lorraine.fr}
	\affiliation{Laboratory of Energetics, Theoretical and Applied Mechanics (LEMTA), URA CNRS 7563,
		University of Lorraine, 54518 Vandoeuvre‐lès‐Nancy, France.}
	\begin{abstract}
		We revisit the optimal performance of a thermoelectric generator
		within the endoreversible approximation, while imposing
		a finite physical dimensions constraint (FPDC) in the form of a 
fixed total area of the heat exchangers. Our analysis 
is based on  the 
linear-irreversible law for heat transfer between the reservoir and the 
working 
medium, in contrast to Newton's law usually assumed in literature.
The optimization of power output is performed with respect to the thermoelectric
current as well as the fractional area of the heat exchangers. 
We describe two alternate designs for allocating optimal 
areas to the heat exchangers. 
		Interestingly, for each design, the use of linear-irreversible law 
yields the efficiency at maximum power in the well-known form, $2\eta_{\rm 
C}^{}/ (4-\eta_{\rm C}^{})$, earlier obtained 
for the case of thermoelectric generator under exoreversible 
approximation, i.e. assuming only the internal irreversibility due to Joule 
heating. On the other hand, the use of Newton's law yields Curzon-Ahlborn 
efficiency.
	\end{abstract}
	\maketitle
\section{Introduction}
	The real-world energy convertors perform under finite-size and finite-time 
	constraints on the resources.
	In recent years, finite-time thermodynamics  \cite{Berry1984} has been 
popular in the study of 
irreversible processes. Finite physical dimensions thermodynamics (FPDT) is 
another approach, which considers, for example, the physical size of heat 
exchanger between heat reservoir and working substance, to study irreversible 
processes in actual devices.
	This approach was started by Chambadal \cite{Chambadal1957} in 1957, 
followed by Novikov \cite{Novikov1958} and further illustrated by   
other authors \cite{Bejan1995A,BEJAN1995,Feidt2017}. For 
instance, Chambadal and Novikov started with a steady-state heat engine which 
is simultaneously in contact with hot and cold reservoirs. It was coupled to 
the hot reservoir through a finite heat transfer conductance and in  
perfect contact with the cold reservoir. Its efficiency at maximum power (EMP) 
comes out in the now well-known form, known as Curzon-Ahlborn (CA) 
efficiency:
	\be
	\eta_{\rm CA}^{} = 1-\sqrt{\theta},
	\ee
	where $\theta =T_c/T_h$ is the ratio of  cold to hot bath temperatures. 
 This EMP is independent of any other model parameters like the Carnot efficiency $\eta_{C}= 1-\theta$. 
 The exact efficiency was reproduced in an elegant way by 
assuming the so-called endoreversible 
approximation where working substance is internally reversible  \cite{CA1975, 
Hoffmann1997}
and the only irreversibilities arise due to non-ideal contacts with the heat 
reservoirs. 

	In this work, we focus on the steady-state energy convertor, working on the
principle of thermoelectricity, which is a paradigmatic model to study the 
effect on performance due to different sources of irreversibility 
\cite{majumdar2004}. We find the optimal power output of thermoelectric generator 
(TEG) in the presence of finite physical dimensions constraint (FPDC).
Here, in addition to optimizing the power 
output with respect to electric current, we also optimize  with respect 
to the fractional area of a heat exchanger. With this 
	step, it will be shown that  the maximum  power output should be at a 
proper selection of the area of the heat exchangers, 
	in addition to an optimal value of the electric current. 
	This selection is an important step in thermal
	optimization, as finiteness of the
	total heat transfer area
	is a relevant constraint
	in the overall design of the energy converter \cite{Bejan1995A}.  
	
		Another objective of this study is to examine the effect of  heat transfer 
law between the working substance and reservoirs on the performance of 
thermoelectric generator. In particular, we investigate the endoreversible
model based on linear-irreversible law of heat transfer. The results are
compared with the usual results based on Newton's law of
heat transfer. 

This paper is organized as follows: In Section II, we describe the Constant 
Properties model of thermoelectric generator along with  
the finite physical dimensions constraint. In Section 
III, power optimization is performed using two different heat transfer laws; 
the variables to optimize are the electric current and  the 
fractional area of the heat exchangers. In Section IV, we discuss 
an alternate design to constrain the areas of the heat exchangers
and discuss its optimal properties. Section V is devoted to a discussion
of the results. We end the paper with Section VI, presenting our conclusions.   
 
	\begin{figure}
	\includegraphics[width=5cm]{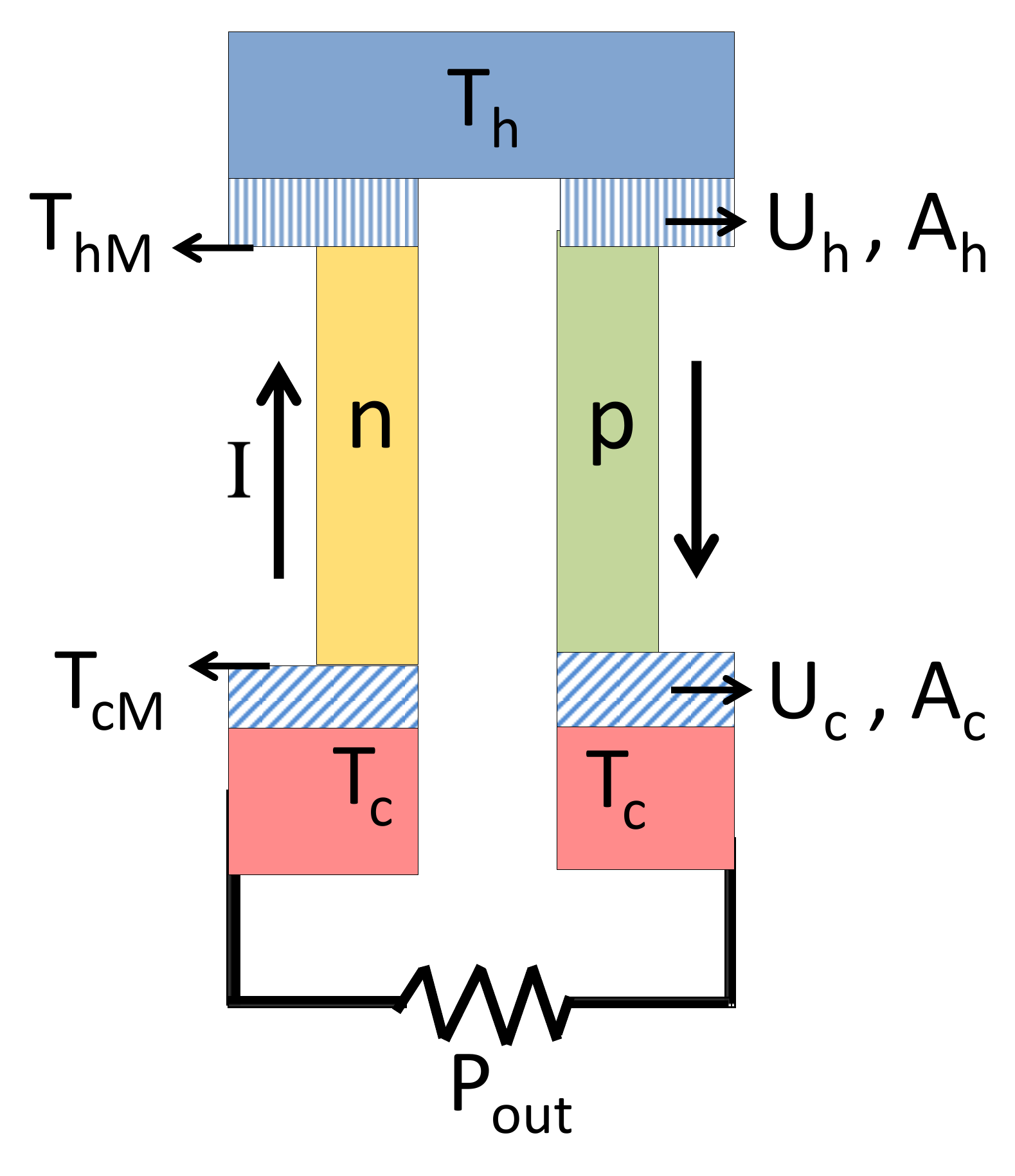} 
	\caption{Schematic of a TEG
		consisting of two legs of thermoelectric material of $n$- and 
		$p$-type, which are connected electrically in series and  
		thermally in parallel. $T_{hM} (T_{cM})$ is 
		local temperature of thermoelectric material on the hot (cold) end.
		In the endoreversible approximation, only the irreversibility
		due to non-ideal thermal contacts with the reservoirs are considered.
		$U_h$ and $U_c$ are the heat transfer coefficients of heat exchangers,  
		with areas $A_h$ and $A_c$ on the 
		hot and cold sides, respectively.}
	\label{fig1}
\end{figure}
	\section{Thermoelectric generator model}
	Thermoelectricity is a non-equilibrium phenomenon, studied within
	the framework of Onsager-Callen
	theory \cite{Onsager1931, Callen1948}. The various thermoelectric effects emerge from the coupling
	between the gradients of temperature
	and electrochemical potential. Within Constant Properties model 
(CPM)\cite{Ioffe1957}, the thermoelectric material 
(TEM) is considered 
	to be a one-dimensional, homogeneous
	substance of length $L$, with given values of
	internal
	resistance $R$, heat transfer conductance $K$, and Seebeck coefficient
	$\alpha$. Further,
	let $I$ denote the constant value of electric current flowing
	through the TEM (see Fig. 1). 
	Then, on the basis of Onsager formalism
	and Domenicali's heat equation \cite{Domenicali1954, Pedersen2007},
	thermal currents 
	at 
	the end points of TEM 
	are written as follows. 
	\bea
	\dot{Q}_h^{} &=& \alpha T_{hM} I +K (T_{hM}-T_{cM})-\frac{1}{2}RI^{2}, 
	\label{hflux}\\
	\dot{Q}_c^{} &=& \alpha T_{cM} I +K(T_{hM}-T_{cM}) + \frac{1}{2} RI^{2} \label{cflux}.
	\eea
	In the above equations, the first term corresponds to 
	convective
	heat flow, where $T_{hM}<T_h$ $(T_{cM}> T_c)$ is the local temperature of
	TEM 
	at hot (cold) end. The second term takes into account heat leakage
	across the TEM, 
	and the last term is the fraction of Joule heat
received by each reservoir, which is equally distributed in case of CPM (see 
also  \cite{Jasleen2019}). Since, 
we are mainly interested in the   
efficiency at maximum power, we shall ignore the parasitic 
heat leaks which reduce the efficiency and consider only the so-called  
strong-coupling assumption ($K \approx 0$)  \cite{Kedem1965}.  
The thermal currents are modified as follows:
	\bea
	\dot{Q}_h^{} &=& \alpha T_{hM} I -\frac{1}{2}RI^{2}, 
	\label{hflux2r}\\
	\dot{Q}_c^{} &=& \alpha T_{cM} I + \frac{1}{2} RI^{2} .
	\eea 
	There are two further limiting operations of a TEG. 
	In the so-called endoreversible approximation, only external 
irreversibility 
due to finite rate of heat exchange between reservoir and TEM is considered. 
	Thus, setting $R = 0$ (when there is no Joule 
heating), thermal currents are written as
	\bea
	\dot{Q}_h^{} &=& \alpha T_{hM} I,
	\label{hflux2}\\
	\dot{Q}_c^{} &=& \alpha T_{cM} I.
	\label{cflux2}
	\eea
In the following, we consider  the problem of optimization of power output 
within endoreversible approximation which is given by the following
condition:
\bea
\frac{\dot{Q}_h^{}}{ T_{hM}} &=& \frac{\dot{Q}_c^{}}{ T_{cM}}, 
\label{endo}
\eea
Due to 
 Eqs. (\ref{hflux2}) and (\ref{cflux2}), each term in the above equation 
 is equal to $\alpha I$. 
Thus, the rate of entropy injection
 at the hot end of TEM and the rate of its removal
 at the cold end of the material are the same, implying that the process
 of energy conversion is assumed to be reversible.
 
We model the flow of heat between a reservoir and the TEM
through the heat exchanger. 
Let $f(T_i,T_{iM})$ represent a general form of the heat 
transfer law, whereby the heat flux through the heat exchanger is given by
\be
\dot{Q}_i = K_i f(T_i, T_{iM}),
\ee
where $i = h, c$ and $K_i$ is the generalized thermal conductance 
of the heat exchanger at the hot or cold end, defined as the 
product of the heat transfer coefficient ($U_i$)
and the area of heat exchanger ($A_i$),  i.e. $  K_i = U_i A_i$. 
Under FPDT, finite dimensions of, say, 
heat exchangers are recognized as optimizable variables 
\cite{Chambadal1957,Novikov1958,BEJAN1995, feidt2012, 
Dong2012,Velasco1997,Lu2018}
in the presence of finite rates of heat transfer.
Thus, the total heat transfer area to be allocated on the hot and cold sides of 
the energy conversion system is constrained: $A_T = A_h + A_c$. 
The performance of TEG will be additionally optimized subject to a given total 
area of the heat exchangers.\
\section{Power optimization}
\subsection{Step 1: Optimization over the electric current}
In the following, we perform the analysis using the heat 
transfer law based on linear-irreversible framework.
Usually, in literature, Newton's law for heat transfer is employed for 
simplicity and analytic solution. As we will see, the present model is also 
exactly solvable.
	\begin{figure}
	\includegraphics[width=0.45\textwidth]{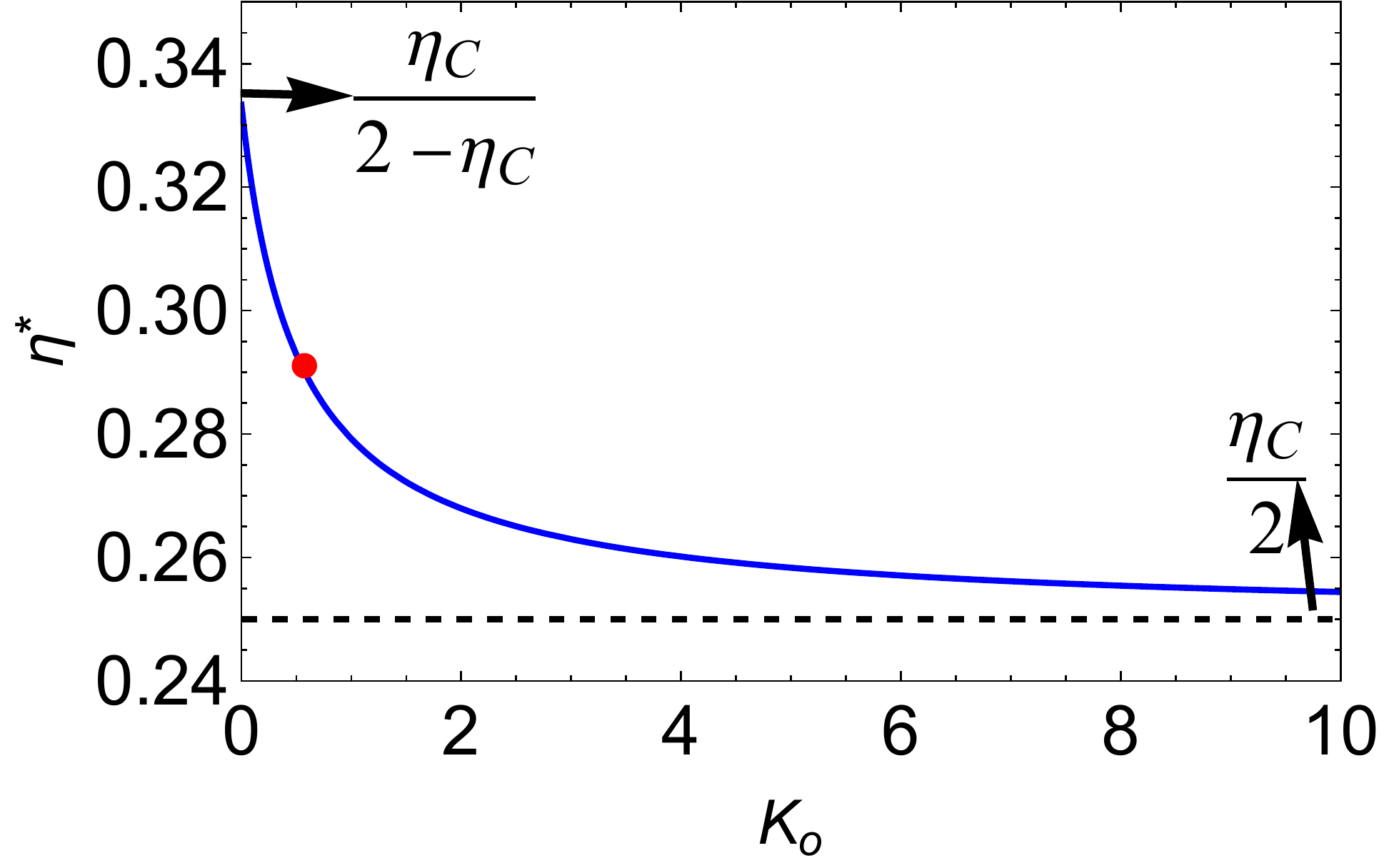}
	\caption{Efficiency at maximum power versus  
		$K_{o}$ = $K_{c}/K_{h}$, Eq.(\ref{Exp2}), for $\theta = 0.5$.
		The upper and lower bounds 
		(marked with arrows) of EMP are obtained for $K_{o}$ $\to$ 0 and $K_{o}$ $\to$ 
		$\infty$ respectively. The dot on the curve depicts the CA-value, 
obtained with $K_o = \theta$.}
	\label{plot2}
	
\end{figure}
According to the linear-irreversible law for heat transfer, the 
heat flux at a thermal contact is proportional to the difference of the 
inverse temperatures
between, say, the reservoir and the working medium. 
The heat flux entering the TEM is thereby given as:
 \be
\dot{Q}_{h}^{} =
K_h^{}\left(\frac{1}{T_{hM}}-\frac{1}{T_h} \right),
\ee
where $K_h \equiv U_h A_h$, with $U_h$ as the heat transfer coefficient
for the heat exchanger based on area  $A_h$. 
$U_h$ is assumed to be independent of temperature. Matching  fluxes at the 
hot interface using Eq. (\ref{hflux2}), the 
hot flux can be written as
\bea
\dot{Q}_{h}^{} &=& \frac{\sqrt{K_h^{2}+4\alpha K_h^{} I T_h^2}-K_h^{}}{2T_h}.
\label{LPhflux}
\eea
	Similarly,  the heat flux entering the cold reservoir is given by: 
\bea
\dot{Q}_{c}^{} &=& K_c^{} \left(\frac{1}{T_{c}}-\frac{1}{T_{cM}} \right),
\eea
and the use of Eq. (\ref{cflux2}) leads to the expression:
\bea
\dot{Q}_{c}^{}	&=& \frac{K_c^{}-\sqrt{K_c^{2}-4\alpha K_c^{} I 
T_{c}^2}}{2T_c}.
\label{LPcflux}
\eea
 
	In the first step towards optimization of power output, $P = 
\dot{Q}_{h}^{}-\dot{Q}_{c}^{}$, upon setting ${\partial P}/{\partial I}=0$, we 
get 
	\be
	I^* = \frac{K_h K_c}{4\alpha (K_h+K_c)}\frac{(T_h^2-T_c^2)}{ T_c^2 
T_h^2}.
	\ee
	The maximum power output, $P^* \equiv P(I^*)$,  is given by
	\be
	P^{*} = \frac{ K_h }{2T_c} \left( {\sqrt{(K_o^{}+1)(K_o^{} + 
\theta^2)}-( K_o^{}+\theta)} \right),
	\label{P*}
	\ee
	where $K_o = K_c^{}/K_h^{}$. Then, the EMP is evaluated to be 
	\bea
	\eta^{*} &=& 1-\frac{\sqrt{(K_o+1)(K_o+\theta^2)}+ \theta-K_o}{1+\theta}.
	\label{Exp2}
	\eea
For a given value of $\theta$, EMP is a monotonically decreasing function of 
$K_o$, 
as depicted in Fig. \ref{plot2}. In particular, EMP is bounded between
two limiting values. For, 
$K_c^{} << K_h^{}$, or, in the limit $K_o \to 0$, we have $\eta^* \to 
\eta_{\rm C}^{}/(2-\eta_{\rm C}^{})$. In the opposite limit, when $K_o\to 
\infty$,  $\eta^* \to  \eta_{\rm C}^{}/2$.
Interestingly, for $K_o=\theta$, the form of EMP is simplified to $ \eta_{\rm 
CA}$. Further, the series expansion of the above EMP for small temperature 
differences, or $\eta_{\rm C}^{}<<1$, is given by:
	\be
	\eta^*= \frac{\eta_{\rm C}^{}}{2}+\frac{\eta_{\rm C}^{2}}{4(1+K_o)}+
	O[\eta_{\rm C}^{3}].
	\label{seriesko}
	\ee
The above series, for $K_o =1$, is given by: $\eta^* \sim {\eta_{\rm C}^{}}/{2} 
+ {\eta_{\rm C}^{2}}/{8}+...$, which 
shows the same universality up to second order that is found for
	strong-coupling heat engines having a left-right symmetry 
\cite{Esposito2009}.

\subsection{Step 2: Optimization over the area constraint}
	Now, the ratio $K_o = K_c/K_h \equiv (U_c/U_h)(A_c/A_h)$ suggests that 
the parameter 
$K_o$ may be tuned by choosing materials with different ratios 
of heat transfer coefficients $(U_c/U_h)$, or by varying the allocation of 
areas $(A_c/A_h)$. Thus, for the given set of materials (fixed  
$U_c/U_h$), there may be a constraint of a fixed total area
to be allocated to the heat exchangers. This constitutes an example of the 
finite physical dimensions constraint (FPDC) mentioned earlier, which 
we analyze in the following.
    
    It is convenient to define the ratios
	$u=U_c/U_h$ and $x=A_h/A_T$. 
	Note that $x$ is the fraction of the total area allocated to the heat 
exchanger at the hot end.
	The maximum power output, Eq. (\ref{P*}), can then be 
written in a dimensionless form as: 
	\begin{align}
\mathcal {P} (x) \equiv   \frac{2T_c}{ U_h A_T} P^*=& 
\sqrt{\{(1-x)u +x \}\{(1-x)u +x \theta^2\}} \nonumber \\
& - \{(1-x)u +x \theta\}.
		\end{align}
	In the second step, we optimize the power output  with respect to 
$x$, for a given value of $u$ and the total area $A_T$.  The optimal 
fraction of the area is found to be
		\be
		\hat{x} = \frac{\sqrt{u}(1+\theta)+2u}{2(1+\sqrt{u})(\theta+\sqrt{u})}.
		\ee
The relative fraction of optimal areas is depicted in Fig. 3.
		The doubly-optimized power, $\hat{\mathcal{P}} = 
\mathcal{P}(\hat{x})$, is
		\be
		\hat{\mathcal{P}} =  
\frac{u(1-\theta)^2}{2(1+\sqrt{u})(1+\theta)}.
		\ee 

    The corresponding EMP is evaluated to be
		\be
		\hat{\eta}=
\frac{\eta_{C}}{2-\gamma \eta_{C}},
		\ee
		where $\gamma = ({1+\sqrt{u}})^{-1}$,
		which has been obtained in different scenarios \cite{Chen1989, 
SchmiedlSeifert2008, Johalepl2018, JohalRai2021, Broeckepl2013}.
For $u \to 0$, the EMP reaches the upper bound discussed earlier and the 
optimal fraction of area on the hot side follows 
$\hat{x} \to 0$. On the other hand,   for $u \to \infty$, the EMP reaches the 
lower bound and the optimal fraction of area on the hot side follows $\hat{x} 
\to 1$.

	\begin{figure}
		\includegraphics[width=0.45\textwidth]{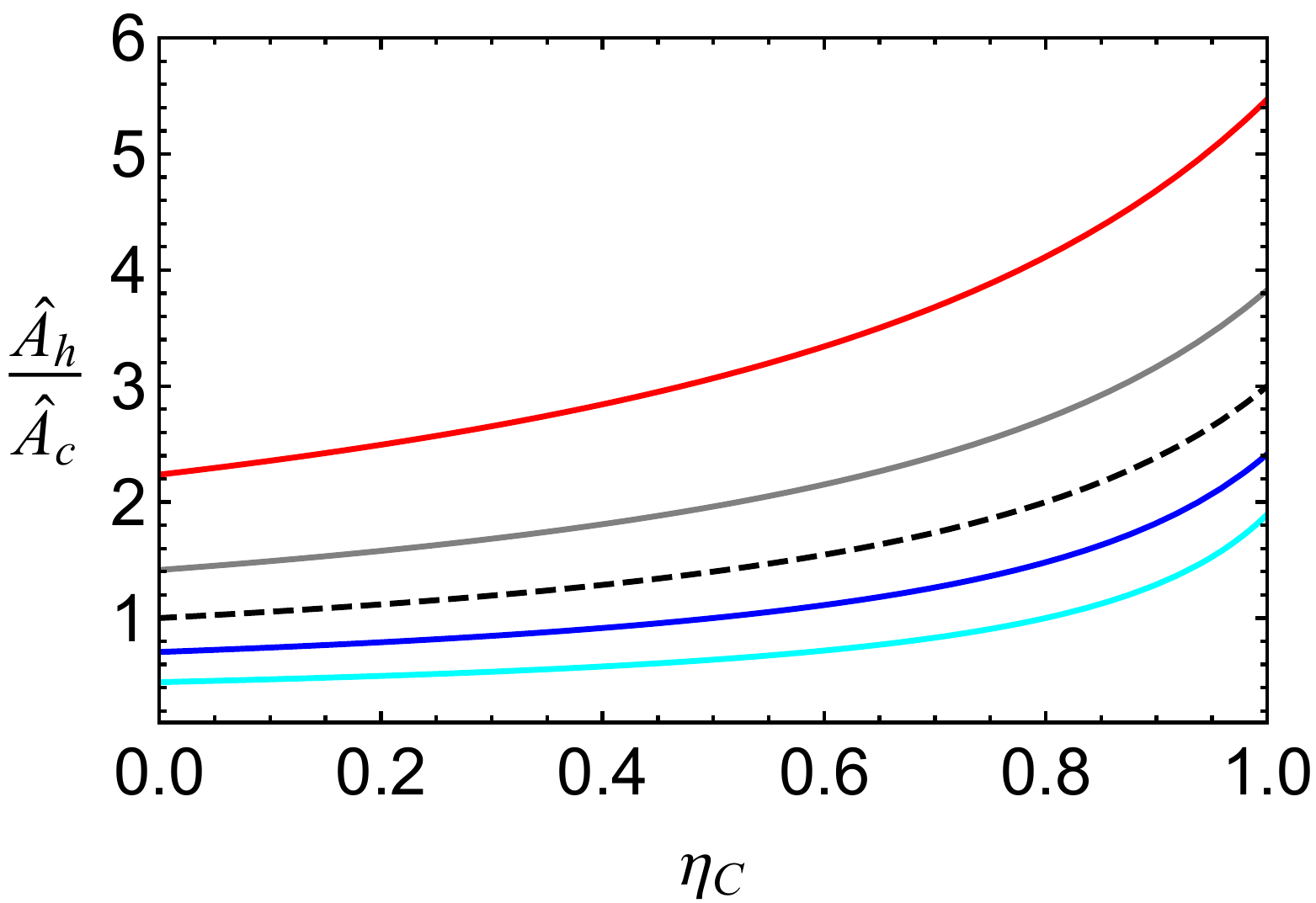}
\caption{Ratio of optimal areas of hot to cold heat exchangers, 
$\hat{A}_{h}/\hat{A}_c = \hat{x}/(1-\hat{x})$,  at maximum power 
output versus $\eta_{C}$ with linear-irreversible heat transfer law, for 
various  $u=U_c/U_h$ values, from bottom to top as 0.2, 0.5, 1, 2, 5. }
\label{plot3a}
	\end{figure}

For $u=1$, the heat exchanger of the same material is to be 
used on the hot and cold sides. The EMP is then simplified to
\be
\hat{\eta} = \frac{2\eta_{\rm C}^{}}{4-\eta_{\rm C}^{}}. 
\label{eta}
\ee 
The above expression also exhibits the universality up to 
second order, as mentioned below Eq. (\ref{seriesko}). 
Here, upon the second step of power optimization, the 'left-right' symmetry 
manifests via the equality of
heat transfer coefficients ($U_h = U_c$) on the hot and cold sides.
However,  the corresponding optimal ratio of areas is  
given as: $(\hat{A}_{h}/\hat{A}_c 
)_{u=1} =(3+\theta)/ (1+3 \theta)$, which implies that $\hat{K}_o = 
(1+3\theta)/ (3+ \theta)$. 
The foregoing case makes it apparant that the second-order universality 
of EMP may be manifested by more general choices of $\hat{K}_o$,  and not
simply for $K_o=1$, as mentioned in Section III.A. 

\subsection{Comparison with Newton's law}
 Next, we employ Newton's law for the finite rate of heat transfer between 
TEM and
heat reservoirs, such that 
\bea
\dot{Q}_{h} &=& K_{h}^{'}(T_{h}^{}-T_{hM}^{}), 
\label{qhnew}\\
\dot{Q}_{c} & =& K_{c}^{'}(T_{cM}^{}-T_{c}^{}),
\label{qcnew}
\eea
where the thermal conductance $K_{i}^{'} \equiv U_{i}^{'} A_i$ and 
$U_{i}^{'}$ is the corresponding heat transfer coefficient. 
Then, applying the flux-matching condition on both hot and cold sides of 
TEM, 
we obtain explicit expressions of the thermal currents
\begin{equation}
	\dot{Q}_{h}^{}= \frac{\alpha T_{h}^{}K_{h}^{'}I}{K_{h}^{'}+\alpha I}, 
	\label{eq6}
\end{equation}
\begin{equation}
	\dot{Q}_{c}^{}=  \frac{\alpha T_{c}^{}K_{c}^{'}I}{K_{c}^{'}-\alpha I}.
\end{equation}
 Optimizing the power output with respect to $I$, the optimal  current 
is
\be
I^*=\frac{K_{h}^{'}K_{c}^{'}}{\alpha(K_{h}^{'}+\sqrt{\theta } K_{c}^{'})} 
(1-\sqrt{\theta}).
\ee
The optimal power output is given by
\be
P^{*} =\frac{K_{h}^{'}K_{c}^{'} 
T_h}{(K_{h}^{'}+K_{c}^{'})}\left(1-\sqrt{\theta}\right)^2,
\label{P_max1}
\ee
and the corresponding hot flux is
\be
Q^{*}_{h} =\frac{K_{h}^{'}K_{c}^{'} 
T_h}{(K_{h}^{'}+K_{c}^{'})}\left(1-\sqrt{\theta}\right).
\label{Q_max1}
\ee
Thereby, the EMP is equal to $\eta_{\rm CA}$.
So, when the  power output is optimized with respect to $I$ using Newton's law,
the EMP is independent of the heat transfer conductances.

In the next step, we incorporate the finite physical dimensions constraint 
in the form of a fixed total area $A_T$, and rewrite the power output,
Eq. (\ref{P_max1}), as
\be
\mathcal{P} \equiv \frac{P^*}{A_T T_h U_{h}^{'}} = \frac{x(1-x) u' 
}{\{(1-x) u'+x 
\}}\left(1-\sqrt{\theta}\right)^2,
\label{Double_Ps}
\ee
where $u'=U'_c/U'_h$.
The power output may be further optimized with respect to $x$, obtaining
the optimum at $\hat{x} =\sqrt{u'}/(1+\sqrt{u'})$.
The doubly-optimized power is given by:
\be
\hat{\mathcal{P}}= \frac{ u' 
}{(1+\sqrt{u'})^2}\left(1-\sqrt{\theta}\right)^2.
\label{Double_P}
\ee

Thus, even though the power
can be doubly optimized while using Newton's law, the EMP does not 
change upon the inclusion of the finite physical dimensions constraint.
 \\

\begin{figure}
	\includegraphics[width=3.5cm]{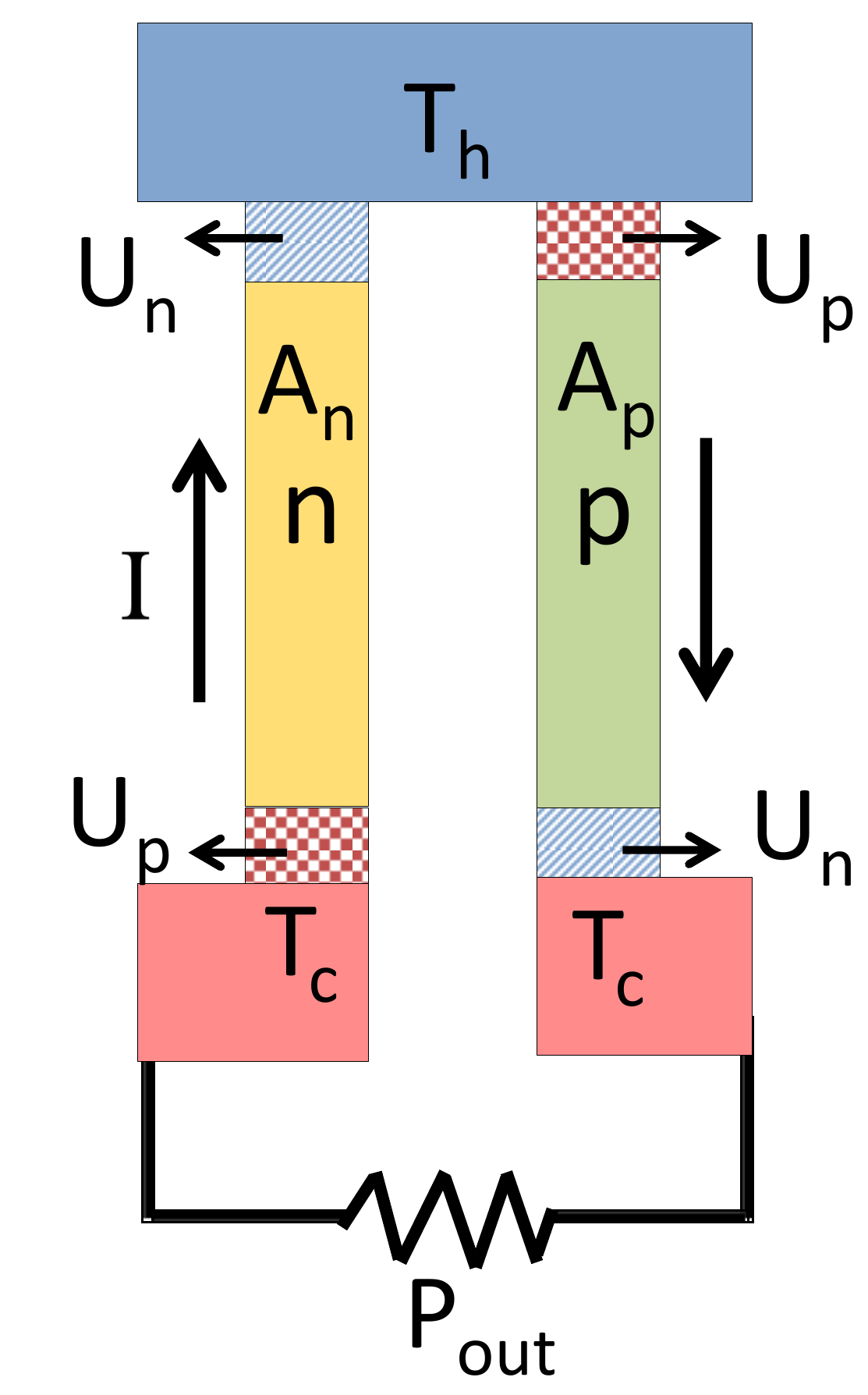}
	\caption{Schematic of a two-leg endoreversible thermoelectric engine when 
the areas of 
		the heat exchangers are matched with the areas of cross-sections of the legs. 
		The  total area on the hot or cold side is fixed to be 
		the total area of cross section of TEM: $A 
= A_n + A_p$. The power output is optimized w.r.t $I$ and $y=A_p/A$, yielding 
the EMP as $2\eta_{\rm C}^{}/ (4-\eta_{\rm C}^{})$.}
	\label{fig4}
\end{figure}
\section{An alternate design}
In the above, the  total area $A_T$ is arbitrary, which may be decided from
the cost of materials, or alternately, from the design constraint. 
As a case study, we analyze a design for the heat exchangers 
based on the two-leg configuration of TEG. 
The areas of cross-section of the $n$-type and $p$-type legs can be $A_n$ and 
$A_p$ respectively \cite{Pedersen2007}. 
The area of a heat exchanger on each (hot or cold)
side is set equal to the area of cross-section of the leg of TEM (see Fig. 
\ref{fig4}), yielding 
the total available area on each side  as 
$A = A_n + A_p$. Further, given two kinds of the heat exchanger
materials with heat transfer coefficients as $U_n$ and 
$U_p$, the materials are distributed as shown in Fig. \ref{fig4}. 
Then, the overall heat conductances on the hot and cold sides are given by:
\be
\begin{aligned}
K_h &= U_p A_p + U_n A_n, \\
K_c &= U_n A_p + U_p A_n,
\label{knkp}
\end{aligned}
\ee
respectively. Now, the first step of power optimization w.r.t $I$
remains the same as discussed in Section III.A. 
Including the area constraint of the present design,
we can rewrite the power at optimal $I^*$, Eq. (\ref{P*}), as
\begin{align}
\mathcal{P}(y) \equiv \frac{2 T_c P^{*}}{A U_n} 
= & \sqrt{(1-\theta^2)(1-v^2)y+(1+v)(v+\theta^2)} \nonumber \\
 & -\{y(1-v)(1-\theta)+v+\theta\},
\end{align}
where $v= U_p/U_n$ and $y=A_p/A$.

Then, in the second step, the above power output is optimized w.r.t $y$, 
obtaining the optimum at
\be
\hat{y} = \frac{3(v-\theta)+v \theta-1}{4(v-1)(1+\theta)}.
\label{opty}
\ee 
Now, since $y$ represents a fraction of the area, we must have
$0\leq \hat{y} \leq 1$. For a given value of $\theta$, this constrains the 
permissible range of $v$ values, as shown in Fig. 5.
There are two regimes:\\
i) For $v<1$, the allowed range of $v$ is
\be
0\leq v \leq \frac{1+3\theta}{3+\theta}    = v_1.
\ee
ii) For $v>1$, the allowed range is
\be
v_2 = \frac{3+\theta}{1+3\theta} \leq v \leq \infty.
\ee 
It implies that in the range $[v_1, v_2]$, there is no 
physically allowed optimal solution of ${y}$, which
also includes the value $v=1$. 
As $\theta \to 1$, this range shrinks and both $v_1$
and $v_2$ approach the value of unity (note that $v_2 = 1/v_1$).

The doubly optimized power output is evaluated as
\be
\hat{\mathcal{P}} =  \frac{(1+v)(1-\theta)^2}{4(1+\theta)}. 
\ee
Remarkably, the EMP for this problem is the same as 
Eq. (\ref{eta}). Also, the optimal value of $K_o = K_c/K_h$, after the above 
optimization,
is given from Eq. (\ref{knkp}) as :
\be
\hat{K}_o = \frac{\hat{y}+v(1-\hat{y})}{v \hat{y} + (1 - \hat{y})}.
\label{optko}
\ee
Upon using Eq. (\ref{opty}) in the above, we get $\hat{K}_o 
= ({1+3\theta})/({3+\theta})$, which is consistent with the findings of Section 
III.B.

Finally, for the case of Newton's law, when the power output is optimized with 
respect to $y$, the optimal point is obtained at $\hat{y}$ = 1/2. 
Thus, 
the optimal areas $A_n$ and $A_p$ come out to be equal at 
the doubly optimized power. The EMP remains at its CA-value.
	\begin{figure}
	\includegraphics[width=8cm]{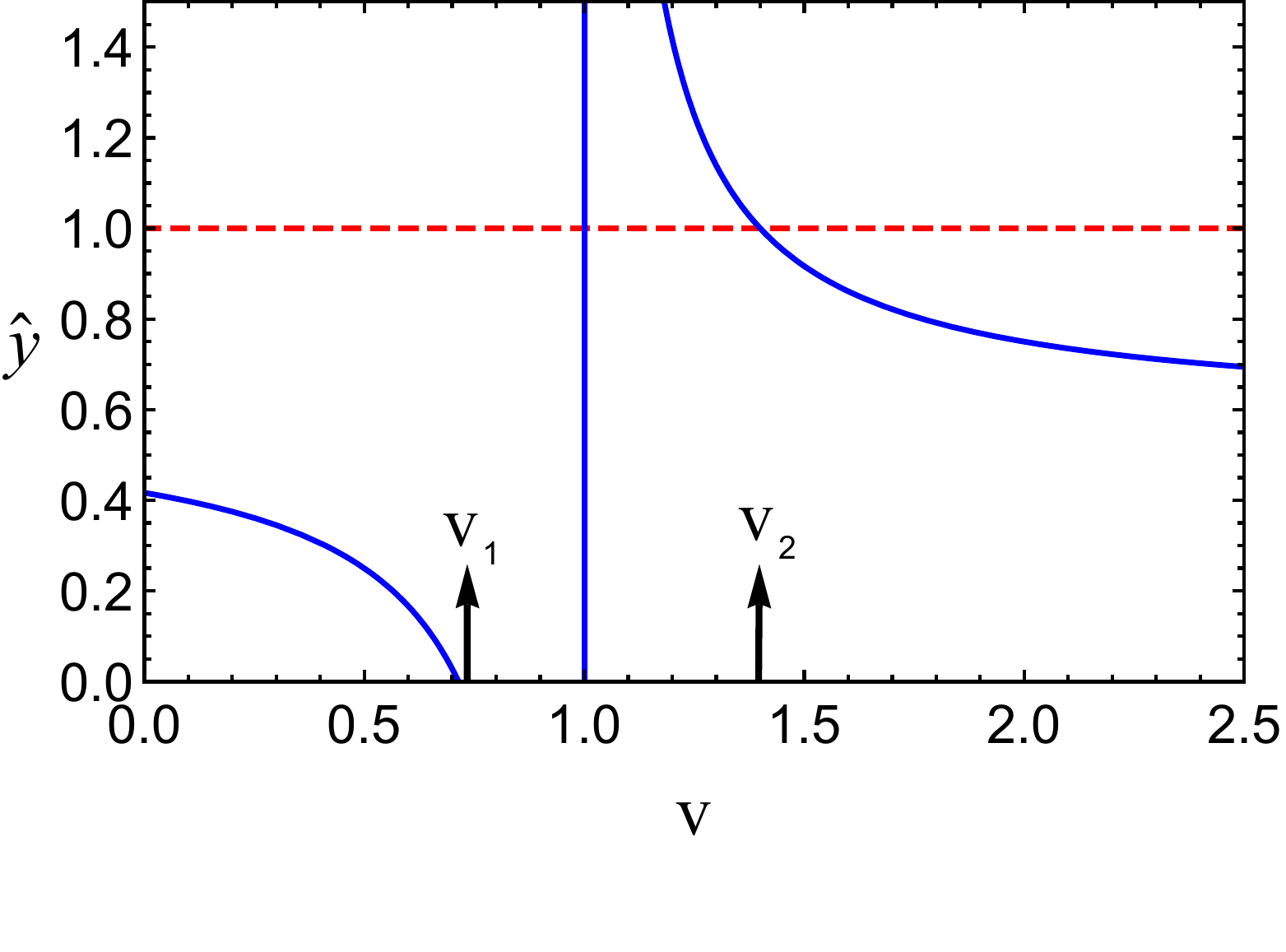} 
	\caption{The optimal solution $\hat{y}$ versus $v$,  Eq. (\ref{opty}), for 
	 $\theta = 0.5$. As the fractional area, the physically
	allowed range $0\leq \hat{y}\leq 1$ yields the corresponding $v$ values
	from $0 \leq v \leq v_1$ and $v_2 \leq v 
\leq \infty$, where $v_2=1/v_1 =(3+\theta)/(1+3\theta)$.}
	\label{plot5}
\end{figure}

	\section{Discussion}
	We have investigated the problem of power optimization in a thermoelectric
	generator where the working medium is modelled within the Constant 
Properties model. As a tractable model, we have focused on the 
endoreversible approximation in the tight-coupling regime. Thereby,
the internal dissipation due to Joule heating and the heat leakage have 
been neglected. 
Usually in literature, Newton's law is employed to model the finite rate
of heat transfer through the heat exchangers. We have investigated the problem 
using the linear-irreversible law based on the difference of inverse 
temperatures.
When the power output is optimized with respect to the electric current, 
a closed form expression for efficiency is obtained (Eq. (\ref{Exp2} )) 
that depends
only on the ratio of thermal conductances of the heat exchangers 
($K_o = K_c/K_h$)
apart from the ratio of reservoir temperatures ($\theta$).
As a second step of the 
optimization strategy, we impose a finite physical dimensions constraint
in terms of a fixed total area of the heat exchangers, given that the 
materials on hot and cold sides can be different. Under this constraint,
we further optimize the power,  which yields an optimal allocation of 
the heat exchanger areas. 
The EMP corresponding to  the doubly optimized power depends
on the ratio of heat transfer coefficients ($u=U_c/U_h$), apart from the ratio
of temperatures. Assuming equal coefficients ($u=1$) on hot and cold
sides, the EMP shows universal features for small temperature 
differences. 

We have also studied an alternate design for the areas of  heat exchangers
based on two materials (with heat transfer coefficients $U_p$ and $U_n$), where 
the total constrained area
is the total area of cross-section of the two legs 
of the thermoelectric module ($A= A_p + A_n$). Interestingly,
the double optimization of power yields 
the EMP which is independent of the heat transfer coefficients.
However, the optimal allocation of areas depends on the ratio $v=U_p/U_n$.

For the purpose of comparison, a similar analysis is performed 
based on Newton's law of heat transfer. The EMP in this case
is the well-known CA value, which is independent of the heat transfer 
coefficients. Here too, the relative areas of the heat exchangers can 
be moved to optimize the power output in the second step.
The optimal areas of heat exchangers are found to be equal in this case. 

In literature, there is an intense discussion on the occurence of universal
expressions of efficiency \cite{Hoffmann1997, Apertet2012B,
Apertet2013A, Apertet2013B, JohalRai2021}.
In the context of thermoelectric 
generators, the exoreversible approximation 
is based on the presence of internal irreversibility ($R\neq 0$)
while assuming ideal thermal contacts with the reservoirs,  
that yields the following relations:
\bea
	\dot{Q}_h^{} &=& \alpha T_{h} I -\frac{1}{2}RI^{2}, \nonumber\\
	\dot{Q}_c^{} &=& \alpha T_{c} I + \frac{1}{2} RI^{2},\nonumber\\
	P  &=& \alpha I (T_{h}-T_{c})  -RI^{2}.
	\eea 
The optimization of power with respect to $I$ yields the EMP
as  $2\eta_{\rm C}^{}/ (4-\eta_{\rm C}^{})$.
 In the present work, we have analyzed endoreversible model
based on the linear-irreversible law for the heat exchange with reservoirs
and performed a double optimization of the power output, first
over $I$ and secondly by imposing the area constraint.
Thus, we come to obtain the same EMP within 
 the endoreversible model as obtained above for the exoreversible model. 
Interestingly, this efficiency is also obtained in discrete endoreversible
heat engines based on linear-irreversible law \cite{Chen1989}.
On the other hand,  the endoreversible model  
using Newton's law yields CA efficiency---with or without
the area constraint.
\section{Conclusions}
	We have considered optimization of the power output of a 
thermoelectric generator based on FPDT, which allows the engineer/designer 
to allocate optimal areas to the heat exchangers,  apart from an
optimal value of thermoelectric electric current.
The approach has been earlier applied to various industrial 
devices, power plants and cooling systems.
The present application to a 
thermoelectric device shows  the utility of FPDT for this class of energy 
conversion devices. In particular, our analysis also highlights the 
comparison between linear-irreversible and Newton's laws  
in thermoelectric engines and provides a toy model to analyze the interplay of 
different forms of the efficiency in these devices.

\end{document}